\documentclass{article}
\usepackage{graphicx}
\begin{document}
\title{Hopping with time-dependent disorder}
\author{G.C.Ferrario\\
Dipartimento di Fisica, Universita' di Milano\\
Via Celoria 16, 20133 Milano,\\
\\
V.G.Benza\\
Facolta' di Scienze, Universita' dell'Insubria\\
Via Lucini 3, 22100 Como\\
also INFM, unita' di Milano} 
\maketitle
\begin{abstract}
We determine the propagation properties of a quantum particle
in a d-dimensional lattice with  hopping disorder, delta correlated in
time. The system is delocalized: the averaged transition probability
shows a diffusive behavior.
Then, superimposed to the disorder, we consider a bias favouring the 
motion with a given orientation, as in the dynamics  of flux lines  
in superconductors.
The result is an effective Liouvillian for the density matrix,
 which is characterized by competition between single particle 
and pair hopping. 
In this case the transition probability is determined in 
terms of excitonic motion, each exciton being extended  along the 
bias direction.
In the small bias regime the hopping disorder is almost uneffective 
 along the Bragg lines of the Brillouin zone, where drift dominates. 
 Elsewere the system undergoes diffusion. 
In the opposite regime we find the single-sided hopping spectrum, 
as  expected from the bias term, but, due to the hopping disorder,
this undergoes an abrupt change of sign at the Bragg lines.

\end{abstract}
\section{Introduction}
Various studies have been devoted to quantum propagation in 
disordered lattices, including  site and hopping disorder.
Here we study time dependent hopping: 
in general the adiabatic motion of a particle in a 
 ``hot'' background.  
  The motion of a charge in a 
rapidly fluctuating effective magnetic field belongs to this 
class: here we give a lattice version of the problem. 
Time-dependent fluctuations of the magnetic field have been considered by  
Aronov and Wolfle in studying the behavior of 
 doped
high-$T_{c}$ materials, close to the metal-insulator transition
\cite{AW}: their analysis was motivated by magnetoresistence 
measurements \cite{JO} in Bi 2:2:0:1 compounds.
Tight binding  hamiltonians were also considered 
for the dynamics of flux lines in superconductors,  
a widely investigated topic, 
both at the experimental and at the theoretical level: see,
e.g.,\cite{G.B.},\cite{V.d.B.} and references therein.
Columnar defects, 
artificially produced by energetic heavy ion radiation, have been 
used in order to pin flux lines and reduce dissipation.
Greatly enhanced pinning has been obtained, e.g., 
 in $YBa_{2}Cu_{3}O_{7}$ crystals with aligned columnar defects,
produced by $Sn$-ion radiation \cite{Civ},\cite{Kon}.
In the corresponding path integral description, the euclidean time is the 
 the vortex line parameter  and 

 the horizontal coordinates of the columnar
defects \cite{NV} define the lattice nodes.
In a hollow cylindrical superconductor the longitudinal current
creates a transverse magnetic field which forces the flux lines 
to tilt with respect to the vertical alignment.
In the hamiltonian this translates into a term, linear in
 the momentum \cite{HN}\cite{Efe}, and antihermitean.
It
can obviously be read as originating from an imaginary vector 
potential. This term explicitly breaks the space inversion 
symmetry: in fact   
the  particle has 
different left and right hopping amplitudes (in a given direction).
 This is not to be confused with a chiral particle,
characterized by single-sided but unitary propagation.
Motion with a preferred orientation arises in various non-quantum 
mechanical contexts,
 e.g. in population
dynamics\cite{NS}, in the transport of  passive scalars in fluids
\cite{MW}, in directed percolation.
The
nonhermitean hopping term has the effect of depinning the vortex
lines, 
as 
shown by various authors:
\cite{HN},\cite{FZ},\cite{E},\cite{Zee}.
Here we first consider 
 time-dependent hopping with no bias. Due to the averaging,
 the appropriate object to be studied,rather than 
the wave function, is the density matrix. In terms of it one 
 reconstructs every transition probability. 
Our approach relies on 
a second quantization formalism, which proved to be very 
efficient in describing edge states in quantum hall systems
\cite{Mathur}.
In the limit of fast fluctuations
 memory effects are canceled and the effective dynamics is  
described by a Liouvillian operator (see Section 2).
 We find that the quantum particle 
 undergoes classical diffusion.
In Section 3 we  add  the deterministic  bias: 
the  Liouvillian, in the second quantization formalism, takes 
then the form of the so-called pair hopping model hamiltonian  
\cite{Sikkema},\cite{Caffarel}.  Pure diffusion is now always frustrated; 
we find excitonic states, propagating with a nontrivial dispersion law.
In Section 4 we summarize our results and compare them with related
work.
In  Appendix 1 we derive a property for averages of 
time-ordered exponential operators; in Appendix 2 we show that
the transition probability, in the hermitean case, 
can be obtained by  resumming the ladder 
diagrams, i.e. that it coincides with the diffuson amplitude.
\section{Disordered lattice}
We start with a lattice hamiltonian with time-dependent hopping
disorder, including an antihermitean term:
\begin{eqnarray}
\label{C1}
{\hat H}_{0}(t)=
&-& \sum_{x,\mu} \big[\big( u(x,\mu;t)+w(x,\mu;t)\big)|x><x+e_{\mu}|\nonumber\\
&+&\big( u^{*}(x,\mu;t)-w^{*}(x,\mu;t)\big)|x+e_{\mu}><x|\big]\nonumber
\end{eqnarray}
here $\mu=1,2,...\,d$, d is the lattice dimension.
We assume zero average gaussian coefficients; the various amplitudes are 
mutually independent, with correlators:  
\begin{eqnarray}
\label{Corr1} 
<u(x,\mu;t)u^{*}(x',\mu';t')>& =& \delta(t-t')\cdot \delta( x 
- x')\cdot \delta_{\mu,\mu'}\cdot D_{u}(x,\mu;t)\nonumber\\
<w(x,\mu;t)w^{*}(x',\mu';t')>& =& \delta(t-t')\cdot \delta( x 
- x')\cdot \delta_{\mu,\mu'}\cdot D_{w}(x,\mu;t)\nonumber\
\end{eqnarray}
As previously announced, we deal with the density operator 
 in the second
quantization formalism. We introduce two, mutually commuting,
fermi (or equivalently bose) operators ${\hat a}(x)$ and ${\hat
b}(x)$,
 related with two independent copies of 
the system. Before averaging, they are 
associated with the retarded and advanced particle , and evolve 
independently with evolutors ${\hat U}$ and ${\hat U}^{*}$.
We define the operator  ${\hat F}$, as the second-quantized 
evolution operator: its matrix elements in the 1+1 particle 
sector are:
\begin{equation}
\label{UU1}
<0|{\hat b} (y) \cdot {\hat a} (x)
\cdot {\hat F}(t,t') \cdot {\hat a}^{+}(x') \cdot {\hat b}^{+}(y')|0>=
<x,y|{\hat U}(t,t')\otimes{\hat U}^{*}(t,t')|x',y'> \nonumber
\end{equation}

${\hat F}$ is  associated with the 
following two-particle hamiltonian:
\begin{eqnarray}
{\hat H}(t)&=&-\sum_{x,\mu}\big[u(x,\mu;t)\cdot{\hat C}(x,\mu)
+ u^{*}(x,\mu;t)\cdot{\hat C}^{+}(x,\mu)\\ 
&+&w(x,\mu;t)\cdot{\hat B}(x,\mu)
- w^{*}(x,\mu;t)\cdot{\hat B}^{+}(x,\mu)\big]\nonumber\\
{\hat C}(x, \mu) &=& {\hat a}^{+}(x)\cdot {\hat a}(x+e_{\mu}) 
- {\hat b}^{+}(x+e_{\mu})\cdot {\hat b}(x)\nonumber\\
{\hat B}(x, \mu)& =& {\hat a}^{+}(x)\cdot {\hat a}(x+e_{\mu}) 
+ {\hat b}^{+}(x+e_{\mu})\cdot {\hat b}(x)\nonumber\\
{\hat Q}(x)&=& {\hat a}^{+}(x)\cdot{\hat a}(x)-
{\hat b}^{+}(x)\cdot{\hat b}(x)\nonumber 
\end{eqnarray}
Using the fact that the disorder is $\delta$-correlated in time, one
can exactly perform the average  of the Neumann series 
and reexponentiate the result, thus obtaining:
\begin{equation}
\label{FF}
<{\hat F}(t,t')>=T exp\big[-sign(t-t')\int^{t}_{t'} d\tau \cdot
{\hat H}_{eff}(\tau)\big]
\end{equation}
The effective Liouvillian ${\hat H}_{eff}$ has the form:
\begin{equation}
\label{Heff}
{\hat H}_{eff}(t)={1 \over 2}\sum_{x,\mu}\big[ D_{u}(x,\mu;t)\cdot
\{{\hat C}(x,\mu),{\hat C}^{+}(x,\mu)\}-D_{w}(x,\mu;t)\cdot
\{{\hat B}(x,\mu),{\hat B}^{+}(x,\mu)\}\big],
\end{equation}
here $\left\{\,,\,\right\}$  denotes the anticommutator.
 The average has generated
a quartic term, which couples the particle and the antiparticle.
The nonlinearity  can be easily handled in this case:  
in fact ${\hat H}_{eff}$ can  be written as 
a quantum spin hamiltonian \cite{Mathur}, if one  starts with the 
fermionic representation. 
 One verifies that an angular momentum algebra is
obtained from ${\hat a}(x)$ and ${\hat b}(x)$.  
The angular momentum is given by:
\begin{eqnarray}
{\hat J}^{+}(x)&=&{\hat a}^{+}(x)\cdot{\hat b}^{+}(x)\\
2\cdot{\hat J}_{3}(x)+1&=& {\hat a}^{+}(x)\cdot{\hat a}(x)+
{\hat b}^{+}(x)\cdot{\hat b}(x)={\hat N}(x)\nonumber
\end{eqnarray}
 If we define $D_{\pm,x,\mu}(t)=D_{u}(x,\mu;t)\pm \cdot D_{w}(x,\mu;t)$;
the Liouvillian turns into:
\begin{eqnarray}
{\hat H}_{eff}(t)&=&-{1 \over 2}\sum_{x,\mu}
\big[4\cdot D_{+,x,\mu}(t)\cdot \big({\hat J}_{1}(x)
\cdot{\hat J}_{1}(x+e_{\mu})
+{\hat J}_{2}(x)\cdot{\hat J}_{2}(x+e_{\mu})\big)\nonumber\\
&+& D_{-,x,\mu}(t)\cdot \big(4\cdot{\hat J}_{3}(x)\cdot{\hat J}_{3}(x+e_{\mu})
+{\hat Q}(x)\cdot{\hat Q}(x+e_{\mu})-1\big)\big]\nonumber
\end{eqnarray}
The planar term, which describes pair hopping,  is  ferromagnetic. 
The vertical term, which  counts the particles, turns from ferro 
to antiferromagnetic as the antihermitean disorder overcomes the hermitean
one.  
The angular momentum operators commute with the charge operators ${\hat Q}(x)$.
Similarly the total number $N_{a}$ of $a$-type
particles, $N_{b}$ , and ${\hat Q}(x)$
 commute with the hamiltonian.
Obviously, as long as we are concerned with the density matrix, 
we are only involved in the $N_{a}=N_{b}=1$ sector, as  
made explicit in 
the matrix elements written in Eq.\ref{UU1}.
In a first class of eigenstates the particles are separated and do 
not propagate, 
since the hopping term   acts only on doubly-occupied sites, i.e. 
on pure states of the form $|x><x|$.
Let us denote such eigenstates as localized. 
A second class  is given by plane wave superpositions 
(magnons) of doubly-occupied sites (in the case of homogeneous disorder): 

\begin{equation}
{\hat J}^{+}(p)|0>\,=\, {1 \over ( 2 \pi)^{d/2}}
\sum_{x}{\hat J}^{+}(x)\cdot exp(i p \cdot x)|0>\,\,(|p_{\mu}| \leq \pi)
\end{equation}
 with eigenvalues :
\begin{equation}
\label{Lapl}
E(p;t)\,=\,2\sum_{\mu}[D_{-,\mu}(t) - D_{+,\mu}(t) \cdot cosp_{\mu}]
\end{equation} 

 The site transition probability is then decomposed in plane wave 
contributions:
\begin{equation}
\label{U2}
\langle|<x|{\hat U}(t,t')|x'>|^{2}\rangle \,=\, 
{1 \over ( 2 \pi)^{d}}\int_{-\pi}^{\pi}dp\cdot
exp\big[ip \cdot(x -x') - sign(t-t')\int _{t'}^{t} d\tau 
\cdot E(p;\tau)\big].
\end{equation} 
 
 The magnons  
are insensitive to any site potential:
 this rather  unintuitive  result  depends on the 
delta-correlation of the hopping amplitudes as shown in Appendix A3.
Hence,
with time-independent correlators  and hermitean
disorder, the Liouvillian reduces to a lattice laplacian: it is
no surprise then that diffusive (long range order) modes 
are the outcome of averaging over fast time-fluctuations. 
If the  disorder has  an antihermitean part,  
  hopping rates locally break space inversion invariance, 
and make some diffusive modes unstable.  
In fact,  
the spectrum of the Liouvillian is no longer positive definite: 
from the minus sign 
in the exponent in Eq. \ref{FF}, one has that
 the portion
of the Brillouin zone inside 
 the   surface  $E(p)\,=\,0$ becomes unstable.
Let us comment on the connection with the problem of a d=2 particle 
in a magnetic field. 
In two dimensions, a fluctuating magnetic 
field orthogonal to the plane would be described by
an hopping coefficient of the form:
 $u(x,\mu,t)= exp (i \cdot
\theta(x,\mu,t))$
with  $\theta(x,\mu,t)$ gaussian; in our model instead  $u$ is  
gaussian.
In spite of this major difference, the present solution  
confirms a previous
result on the motion in a fluctuating magnetic field, obtained 
in the continuum case
 \cite{BC}.
 The effective motion of the quantum
particle is in both cases classical diffusion. 
We are not able to understand wether this is 
 a mere coincidence or is related with some general property 
shared by both approaches.

\section {Biased system with disorder}
We add now a deterministic asymmetric hopping, which describes a 
biased transport in a preferred direction\cite{NS}.
The  hamiltonian, for the particle-antiparticle system, 
 becomes:
\begin{eqnarray}
{\hat H}_{bias}\,&&=\,\sum_{x,\mu}\delta_{\mu,\overline{\mu}} \alpha [exp(-k)
{\hat a}^{+}(x+e_{\mu})\cdot {\hat a}(x)+exp(k)
{\hat a}^{+}(x)\cdot {\hat a}(x+e_{\mu})]\nonumber\\
&&-\sum_{x,\mu}\delta_{\mu,\overline{\mu}} \alpha [exp(-k)
{\hat b}^{+}(x+e_{\mu})\cdot {\hat b}(x)+exp(k)
{\hat b}^{+}(x)\cdot {\hat b}(x+e_{\mu})],
\end{eqnarray}
where $k$ and $\alpha$ are real.
Since in the perturbative series one can isolate the deterministic 
term and expand in the disorder term (see Appendix A1), the total
effective Liouvillian ${\hat L}$ is simply:
${\hat L}= i \cdot sign(t-t')\cdot{\hat H}_{eff}(t) + {\hat H}_{bias}$,where 
 ${\hat H}_{eff}$ is given in Eq. \ref{Heff}.
If we consider homogeneous  hopping disorder,  we recover 
a non-hermitean version
of the so-called pair hopping model hamiltonian \cite{Caffarel}.
Notice that 
here the pair hopping term is intrinsically dissipative.  
Neither the magnons nor the eigenstates of the bias term  
(free-particle states) are  eigenstates. 
It is nonetheless possible to determine two families 
of solutions, which can be regarded as the natural extension 
of the previously determined ones (localized and diffusive, respectively).  
The wave function $f(x,y)$, in the two-particle  
 sector $(N_{a}=N_{b}=1)$, satisfies the eigenvalue equation
\begin{eqnarray}
& \alpha &\cdot[e^{-k}f(x-e_{\overline{\mu}},y)+e^{k}f(x+e_{\overline{\mu}},y)
-
e^{-k}f(x,y-e_{\overline{\mu}})-e^{k}f(x,y+e_{\overline{\mu}})]
\\
+ & i & \sum_{\mu}{D_{+}(\mu) \delta_{x,y}
[f(x+e_{\mu},y+e_{\mu})+f(x-e_{\mu},y-e_{\mu})]-2 D_{-}(\mu)f(x,y)}=\mathcal{E}
\cdot f(x,y),\nonumber
\end{eqnarray}
where, with respect to the previous section's notation, we have 
$\mathcal{E} \equiv -i \cdot E$.
Since the pair-hopping term vanishes  on singly-occupied sites, 
 two-particle states, separated in every other direction
but the bias one, will only be acted by the single hopping term.
This  first class of solutions  is the obvious extension of the 
formerly localized ones.
In the second class  there is no longer
separation orthogonal to the bias.
Let us  proceed to the details of the solution.
Upon writing $f$ as a function of the baricentric coordinate $R=(x+y)/2$ 
and of the relative coordinate $r=x-y$,
one easily identifies the eigenspace $S^{0}$,
which can be spanned in terms of the eigenfunctions:
\begin{equation}
\label{funcE}
f_{\mathcal{E},n,R^{0}}(R,r)=
 exp(iP \cdot R + i q_{\overline{\mu}} \cdot r_{\overline{\mu}})
\cdot [\Pi_{\mu \neq \overline{\mu} }  \delta_{r_{\alpha},n_{\alpha}}\,
\delta_{R_{\mu},R^{0}_{\mu}}],
\end{equation}
where $R^{0},n$ are(d-1)-dimensional vectors,
playing the role of degeneracy indexes.
They are the projection of the baricentric and relative coordinates
 on the space $X_{\perp}$, orthogonal 
to the bias.
 The  eigenvalues then, up to a constant,
 coincide with the eigenvalues of ${\hat H}_{bias}$ :
\begin{equation} 
\label{eigenv}
\mathcal{E}(P_{\overline{\mu}} ,q_{\overline{\mu}}) = 
 -4 \alpha \cdot  sin[(P/2)_{\overline{\mu}} -ik]
 \cdot  sin[ q_{\overline{\mu}}]
-2i \sum_{\mu}D_{-}(\mu).
\end{equation}
 
One verifies that $S^{0}$ is   eigenspace 
also upon adding a static disordered potential $V(x)$.
 Spectrum and eigenfunctions will then reproduce 
the features discussed by 
Hatano and Nelson \cite{HN}.
To go over to the second class of solutions we first
 Fourier transform  the equation:
\begin{eqnarray}
(\mathcal{E} - 2 \alpha \cdot  cos[(P/2 - q)_{\overline{\mu}} -ik]
+ 2 \alpha\cdot  cos[(P/2 + q)_{\overline{\mu}} -ik])f(P,q)=\\
= -2i \sum_{\mu}D_{-}(\mu)f(P,q) +2i \sum_{\mu}D_{+}(\mu)cos (P_{\mu})
{1 \over (2\pi)^{d}} 
\int_{-\pi}^{+\pi} d \overline{q} f(P,\overline{q})\nonumber
\end{eqnarray} 
 
After integrating over $q$, we get:
\begin{eqnarray}
(i/2\pi) \int_{-\pi}^{+\pi}{d q' \over (\eta/z) +
sin(q')}&=&{z \over 4 \cdot \xi}          \\
\eta=(\mathcal{E}/4) +i/2 \cdot \sum_{\mu}D_{-}(\mu);\,&\,&\,z=\alpha \cdot sin
[P_{\overline{\mu}}/2 -ik]\,\,; 
\xi  \equiv {1 \over 2} \sum_{\mu}D_{+}(\mu)cos (P_{\mu}),\nonumber
\end{eqnarray}
where the integral is one-dimensional.
The eigenvalues are: 
\begin{equation}
\label{spectrum}
\mathcal{E}(P)=-2i \cdot\sum_{\mu}D_{-}(\mu) + 4i
\xi \cdot[1-({z \over  \xi})^{2}]^{1/2}
\end{equation}
Notice that 
the solution  is  invariant under the symmetry 
$P\to -P\,\,, k \to -k$; out of the two branches of the square root, 
we take
the one that, as $\alpha \to 0$, goes into the 
spectrum of the unbiased case (Eq. \ref{Lapl}). 
The wavefunction $f_{P_{0}}(P,q)$, associated with
$\mathcal{E}(P_{0})$,  has the form:
\begin{equation}
\label{Ffunction}
f_{P_{0}}(P,q)={{i \xi
\cdot \delta (P - P_{0})} \over 
{ z \cdot sin(q_{\overline{\mu}}) +i\xi \cdot[1-({z \over  \xi})^{2}]^{1/2} }}.
\end{equation} 
In coordinate representation one finds an exponential behavior:
$f \approx \zeta_{\pm}^{r_{\overline{\mu}}} $, where:
\begin{equation}
\zeta_{\pm} = {\xi \over z} \cdot \big[ [1 -({z \over
\xi})^{2}]^{1/2}\pm 1 \big],\,\,(\zeta_{+}\cdot\zeta_{-}= -1) \nonumber
\end{equation} 
and  $\pm$ is to be chosen according with the 
condition $|\zeta_{\pm}|<1$.

 The function $f$ is divergent as $r_{\overline{\mu}} \to -\infty$,
but when computing
 the transition  probability between two sites 
  one only 
needs $r_{\overline{\mu}}=0$ both in $f$ and in the solution 
of the transpose equation.  
We finally examine the case with $\zeta_{\pm}$
lying  on the unit circle.  

 The condition $|\zeta_{\pm}|=1$  implies
$(\xi/z)$  real, with absolute value smaller than one: 
this is verified only in the absence of asymmetry in 
the deterministic term ($ k=0$), or when 
 $P_{\overline{\mu}}= \pm \pi$.
We have  an  exciton, extended along $\mu = \overline{\mu}$ , 
   with global momentum $P_{0}$  
and    relative momentum $p_{eff}$: 
\begin{eqnarray}
 p_{eff}&=&arcos \big({\xi(P_{0}) \over {\alpha \cdot
sin[(P_{0})_{\overline{\mu}}/2]}}\big)\,\, ( k=0),\\
 p_{eff}&=& arcos \big({\xi(P_{0}) \over {\alpha \cdot
cosh[k]}}\big)\,\, (P_{\overline{\mu}}= \pm \pi)\nonumber
\end{eqnarray} 
 The spectrum of the Liouvillian  $E(P)$ (Eq. \ref{spectrum}) in d=2, 
with bias in the
$ \overline{\mu}=1$ direction, and with
 hermitean, isotropic 
disorder
( $D_{w}(\mu)=0$, $D_{u}(\mu)=D$),   
is exhibited in the Figures.
 The imaginary part  $E_{I}(P)$ of $E(P)$ 
describes the reversible motion,
with drift velocity $v = \nabla_{P} E_{I}(P)$.
The real part $E_{R}(P)$ describes the irreversible 
motion: for $k \ne 0$ a region of instability at the center 
of the Brillouin zone appears.
When $\alpha/D$ is small enough, i.e. when disorder dominates,
drift is found along the
 Bragg lines $P_{2}= \pm P_{1} \pm \pi$
(see Fig.1); out of such lines $E_{I}(P)$ is practically zero
and the exciton predominantly undergoes diffusion.
Upon increasing $\alpha$, the bias 
exponent $k$  tilts the 
plane $E_{I}(P)$ in the $P_{1}$ direction, thus enforcing the drift 
also at the center of the Brillouin zone.
 This is exibited in Fig.2  At very strong bias ( $k$ large) one would
expect the single-sided hopping to dominate. The situation 
is rather different (see
Fig.3). 
One 
 easily verifies that $E(P) \approx sign[(cos(p_{1}) + cos(p_{2})]
\cdot (\alpha \cdot exp(k)) \cdot 
(cos(p_{1}/2) +i \cdot sin(p_{1}/2)$, where the  last factor
is indeed the eigenvalue of the bias operator in the single-sided hopping 
limit. The effect of disorder is in the first factor: 
an abrupt change in sign of both the dissipative and reversible 
part of the spectrum ay the Bragg lines.
Let us finally discuss the case $k=0$, which describes a particle with 
deterministic anisotropic diffusion plus disorder. On qualitative
grounds, everything goes as in Fig.1, but now $E_{R}(P)$ is always 
positive (no instability occurs). 
Reversible and irreversible motion are now completely separated, 
in dependence on $P$.
 Around the Bragg lines,  
 we have only drift, apart from a constant damping factor; in the
complementary region 
we find pure diffusion,  with $E_{I}(P)=0$ (no drift). 
 The drift region tends to broaden as $\alpha$ is increased.
\section{Conclusions}
We have discussed the motion  of a particle over a lattice with  
fastly fluctuating hopping amplitudes: the model describes a 
massive quantum object coupled with a high temperature 
background.
In our formalism the transition probability is written as a
transition amplitude for a two-particle quantum system.
This makes simpler the operation of averaging the probability 
over classical fluctuations.
Before averaging, the two particles evolve independently, respectively
forward and backward in time;
after averaging, they interact and their motion becomes irreversible.
  Their effective hamiltonian, which  is simply
the  Liouvillian of the density matrix for the original system, 
has a quartic interaction, which can be readily   
put in the form of  an Heisenberg  hamiltonian.
We determined the steady states of the Liouvillian: with them
 one computes the density matrix and  
any single-particle transition probability (see Eqs.1,8). 
A first class of steady states has the two particles physically separated 
and localized.  The site transition
probability  depends on double occupancy  states: it is  
a sum of plane waves, evolving in time with a diffusive law. 
The generic behavior of  such quantum systems is then diffusion, and
this holds true even in the presence of a  disordered  potential, 
as shown in Appendix 3: since the hopping amplitudes are 
delta-correlated in time, they destroy the phase coherence 
of the wave function. Quantum interference effects, essential 
for localization, are thus absent.
Diffusion in a quantum-mechanical system was found in the Harper model
at its critical point \cite{HA},\cite{GK}. 
The present result, obtained from a lattice model,  confirms 
 a previous one, on the motion 
of a particle 
in a rapidly fluctuating magnetic field, derived in the continuum 
case \cite{BC},.
In Section 3 we added a deterministic, anisotropic bias,
 enforcing a favoured orientation along a given direction; as
already 
illustrated, this term arises quite naturally in 
 describing tilted vortex motion in 
superconductors.
 The Liouvillian takes then  the form of the so-called pair-hopping model
hamiltonian (i.e. it includes both single-particle and pair hopping,
and the two terms do not commute).
In our context the coupling constants are complex, since we are mixing 
reversible and irreversible motion.
 Two classes of steady states can be found.
In the first class   
 the two particles  do not interact, 
provided that their wave packets do not overlap in the plane
orthogonal to the bias.
On such states the hopping disorder has practically no effect, and the
interesting physics is the depinning transition, as described 
by Hatano and Nelson. 
In the second class  the two particles form an exciton, 
 extended along the bias direction.
The dispersion law is a nontrivial function of the exciton momentum.
For small enough bias,  the Brillouin
zone splits into a diffusion-dominated and a drift-dominated part, the
latter lying around the Bragg lines.
The site transition probability, which adds over the 
 the exciton contributions, is the sum of two
parts:  essentially reversible evolution around the
Bragg lines, and irreversible diffusion with no drift
in the complementary region.
Notice that this is different from a mere sum of the two types 
of motion, since the separation involves different regions of 
momenta.
For very large bias  the dispersion law reduces to the one-way hopping
form $ exp(i\cdot p_{1}/2)$, but multiplied by
 the sign of $cos(p_{1})+cos(p_{2})$: this is the signature of the 
hopping disorder, which translates into a singular behavior along the 
Bragg lines.
It is seen then that a perturbative approach fails also in the extremal 
regimes.
\section{Appendix A1}
We first point out here a relevant property of
gaussian averages  of time-ordered exponential operators, 
holding for perturbations delta-correlated in time.
Let us consider the operator ${\hat H}(t)={\hat H}_{0}(t) + {\hat
V}(t)$, where the perturbation term is given through its correlator
$<{\hat V}(t)\otimes {\hat V}(t')>\,=\, \delta(t-t')\cdot {\hat A}(t)$.
One has, by definition:
\begin{displaymath}
Texp\big[\int_{t'}^{t} d \tau \cdot {\hat H}(\tau)\big] 
= \sum_{l=0,p=\pm}^{\infty}p^{l}\int_{t_{l+1}=t'}
\big[ \prod_{m=1}^{l} dt_{m}\cdot \theta(p(t_{m}-t_{m+1}))
\cdot {\hat H}(t_{m})\big]\cdot \theta(p(t-t_{1}))
\end{displaymath}
The average leads to:
\begin{eqnarray}
<Texp\big[\int_{t'}^{t} d \tau \cdot {\hat H}(\tau)\big]>
&=&Texp\big[\int_{t'}^{t} d \tau \cdot [{\hat
H}_{0}(\tau) + sign(t-t')\cdot {\hat V}_{eff}(\tau)]\big]\nonumber\\
<{\hat V}(t)\cdot {\hat V}(t')>&=&2\cdot \delta(t-t')
\cdot {\hat V}_{eff}(t)\nonumber 
\end{eqnarray}
\section{Appendix A2}
The calculation of the averaged transition probability (eq. \ref{U2}),
in the hermitean case, can be performed also by means of an exact 
resummation of ladder diagrams;  the probability then coincides 
with the diffuson amplitude.
The retarded and advanced Green's functions are:
\begin{eqnarray}
[-i \partial_{t} \mp i \eta + {\hat h}(t)]\cdot {\hat G}^{\pm}(t,t') 
&=&\delta(t-t')\nonumber\\
{\hat G}^{\pm}(t,t')&=& \pm i \theta(\pm(t-t'))\cdot {\hat
U}(t,t')\nonumber\\
{\hat G}^{+}(t,t')&=&[{\hat G}^{-}(t',t)]^{+}\nonumber\\
|<x|{\hat U}(t,t')|x'>|^{2}&=&|<x|{\hat G}^{+}(t,t')|x'>|^{2}+
|<x|{\hat G}^{-}(t,t')|x'>|^{2}\nonumber
\end{eqnarray}
The average of the ${\hat U}$ operator is performed first, obtaining 
an effective single particle generator ${\hat h}_{eff}(t)$:
\begin{eqnarray}
<{\hat U}(t,t')>&=&Texp\big[-sign(t-t')\int_{t'}^{t} d \tau
\cdot {\hat h}_{eff}(\tau)\big] \nonumber\\
{\hat h}_{eff}(t)&=& \sum_{\mu,x}D(\mu;t)|x><x|\nonumber
\end{eqnarray}
Let us consider now the perturbative expansion for the average of
$<x|{\hat G}^{+}(t,t')|x'>\cdot <x'|{\hat G}^{-}(t',t)|x>$,
where ${\hat h}(t)$ is the perturbation.
One can show that the expansion  can be written as the sum of
particle-antiparticle diagrams where the 'free propagator' lines are 
substituted with the exact averaged Green's functions 
${\hat G}_{av}^{\pm}$, and the 
contractions are made only between particle and antiparticle 
disorder vertices.
In fact, when contracting particle-particle  vertices one 
is computing contributions to the averaged propagator.
Due to causality and to delta correlation in time, crossed diagrams 
are zero ; one is left with the sum of ladder diagrams, 
i.e. only the diffuson survives. 
The basic contribution to the ladder has the form:
\begin{eqnarray}
&&<t,x;\overline{t},\overline{x}|{\hat M}|t',x';\overline{t'},\overline{x'}>
=\sum_{y,\overline{y}}\langle<x|{\hat h}(t)|y>
\cdot <\overline{y}|{\hat h}(t)|\overline{x}>\rangle_{av}\nonumber\\
&&\cdot <y|{\hat G}^{+}_{av}(t,t')|x'>\cdot
<\overline{x'}|{\hat G}^{-}_{av}(\overline{t'},\overline{t})|\overline{y}>
=\delta(t-\overline{t})\cdot \delta_{x,\overline{x}}\cdot 
\delta_{x',\overline{x'}}\cdot\theta(t-t')\cdot\theta(t-\overline{t'})
\nonumber\\
&&\cdot\sum_{\mu}D(\mu;t)\cdot [\delta_{x,x'-e_{\mu}}+\delta_{x,x'+e_{\mu}}]
\cdot exp\big[(-\sum_{\mu}\big(\int_{t'}^{t} d\tau D(\mu;\tau)
-\int_{\overline{t'}}^{t} d\tau D(\mu;\tau)\big)\big]\nonumber 
\end{eqnarray}
Let us introduce the following operator:
\begin{displaymath}
<t,x|{\hat P}|t',x'>=\theta(t-t')\sum_{\mu}D(\mu;t)\cdot
[\delta_{x,x'-e_{\mu}}+\delta_{x,x'+e_{\mu}}]\cdot exp\big[-2\sum_{\mu}
\int_{t'}^{t} d\tau D(\mu;\tau)\big],
\end{displaymath}
then use the identity:
\begin{displaymath}
<t,x;\overline{t},\overline{x}|{\hat M}^{l}|t',x';\overline{t'},\overline{x'}>
=\delta(t-\overline{t})\cdot \delta_{x,\overline{x}}\cdot
\delta_{x',\overline{x'}}\cdot <t,x|{\hat P}^{l}|t',x'>.
\end{displaymath} 
The diffuson amplitude $\Delta(t,x;t',x')$ is  given by:
\begin{eqnarray}
\Delta(t,x;t',x')&=&\int dt_{1}\theta(t-t_{1})\cdot exp\big[-2\sum_{\mu}
\int_{t_{1}}^{t}d \tau \cdot D(\mu;\tau)\big]<t_{1},x|\sum_{l=0}^{\infty}
{\hat P}^{l}|t',x'>\nonumber\\
&=&\theta(t-t')<x|Texp\big[ \int_{t'}^{t} d\tau \cdot \big({\hat S}(\tau) 
-2\sum_{\mu}D(\mu;\tau)\big)\big]|x'>\nonumber\\
{\hat S}(t)&=&\sum_{x,\mu} D(\mu;t)\cdot 
\big[|x><x+e_{\mu}|+|x+e_{\mu}><x|\big]\nonumber
\end{eqnarray}
In the momentum representation one  recovers  diffusion:
\begin{displaymath}
\Delta(t,x;t',x')={\theta(t-t') \over (2 \pi)^{d}} 
\int_{-\pi}^{+\pi}dk\cdot exp\big[ik\cdot (x-x')-2\sum_{\mu}
 \int_{t'}^{t}d\tau \cdot D(\mu;\tau)\cdot (1-cosk_{\mu})\big]
\end{displaymath}
\section{Appendix A3}
 Let us consider the interaction representation by taking the hopping 
term as a perturbation; we have the evolution operator ${\hat O}$:
\begin{equation}
\label{Gauge1}
{\hat O}(t,t_{0})= \sum_{x} 
exp \big[ -i \int_{t_0}^{t} d \tau \cdot V(x;\tau) \big]|x><x|
\end{equation}  
The transformed hamiltonian ${\hat h}(t)$ is then: 
${\hat O}^{+}(t,t_{0}) \cdot {\hat h}_{0}(t) \cdot 
{\hat O}(t,t_{0})\,=\, {\hat h}(t)$; one verifies that ${\hat h}(t)$ is 
obtained from ${\hat h}_{0}(t)$ through the substitution:
\begin{eqnarray}
\label{Gauge2}
u(x,\mu;t) &\to& u(x,\mu;t)\cdot
exp \big[ i \int_{t_0}^{t} d \tau \cdot \big( V(x+e_{\mu};\tau)
-V(x;\tau)\big) \big]\\
w(x,\mu;t) &\to& w(x,\mu;t)\cdot
exp \big[i \int_{t_0}^{t} d \tau \cdot \big( V(x+e_{\mu};\tau)
-V(x;\tau)\big) \big]\nonumber
\end{eqnarray} 
 The invariance of correlators under this transformation is 
the origin of the independence of Eq. \ref{U2} on the potential.
Notice further that the following identity holds in general:
\begin{equation}
\label{Evol}
{\hat U}_{0}(t,t')={\hat O}(t,t_{0})\cdot {\hat U}(t,t')\cdot
{\hat O}^{+}(t',t_{0}),
\end{equation}
where ${\hat U}_{0}$  and ${\hat U}$ are  the evolution operators 
of the original hamiltonian ${\hat h}_{0}$ (see Eq. \ref{C1}) and
of ${\hat h}$ respectively.
\section{Figure captions}
Fig.1a: real part $Er$ of $E(P)$, over one half of the Brillouin zone 
$(- \pi < P_{1} < \pi\,\,, 0.< P_{2} < \pi)$, for the parameters 
$D=1.,\,\,k=0.5,\,\,\alpha=0.1$\\
Fig.1b: imaginary part $Ei$ of $E(P)$, same parameters as in Fig.1a.\\
Fig.2a,b: competition between diffusion and bias: same as in 
Fig.1, but with $D=1.\,\,,k=0.6\,\,,\alpha=0.75$\\
Fig.3a,b: strong bias:$D=1.\,\,,k=5.\,\,\alpha=0.75$.\\

\begin{table}
\begin{tabular}{c c}
\includegraphics[width=8 cm] {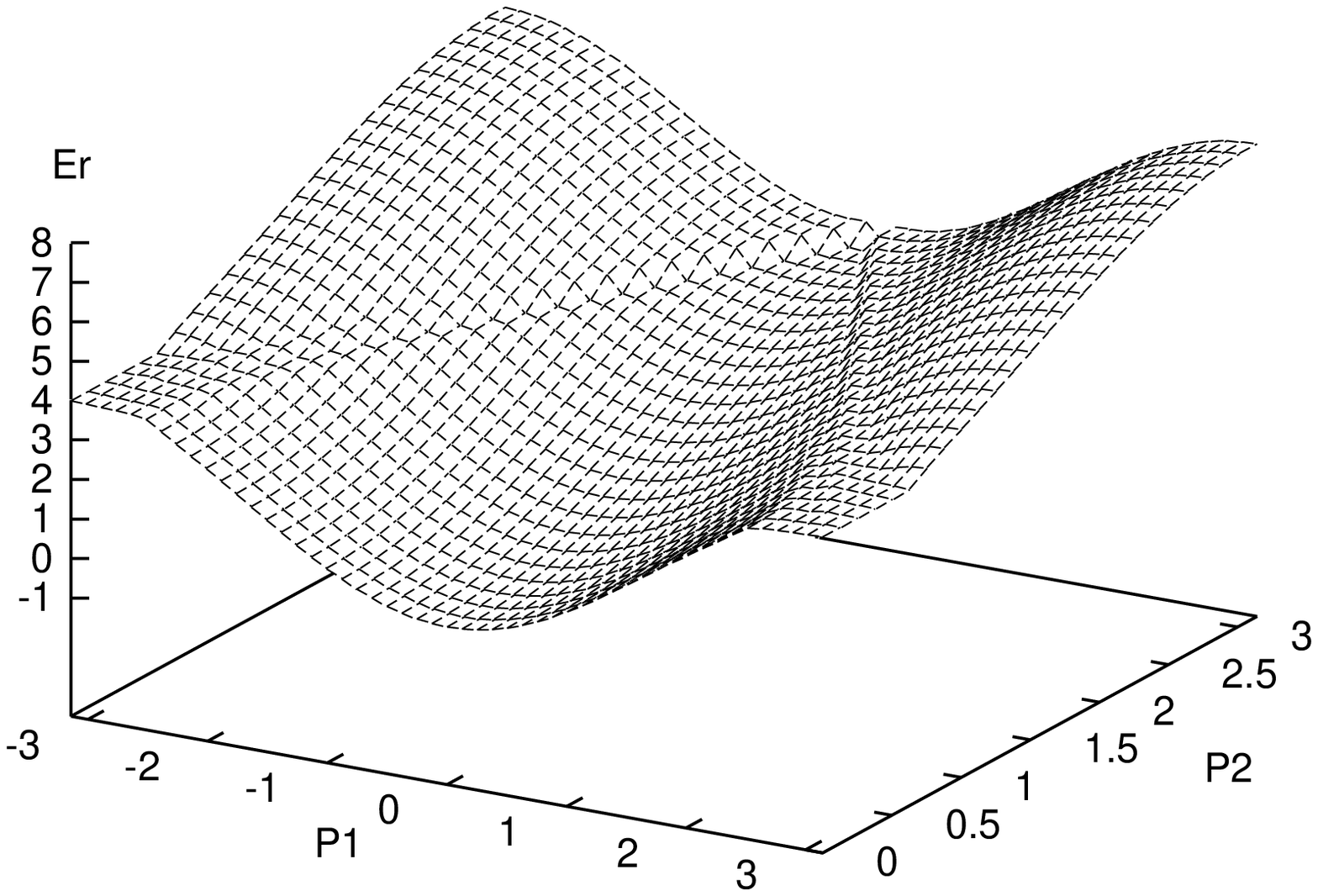}
&\includegraphics[width=8 cm] {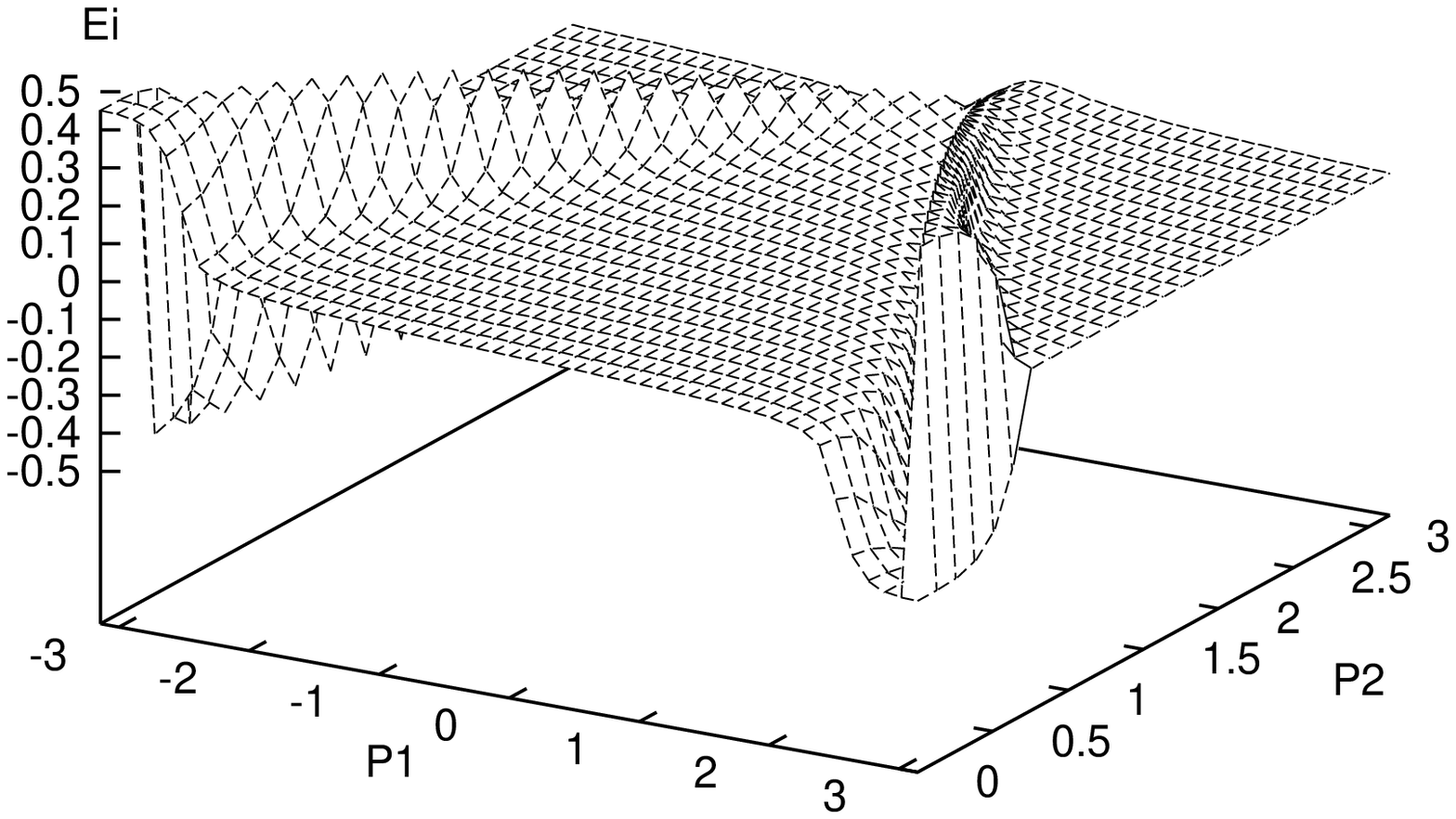}\\
\includegraphics[width=8 cm] {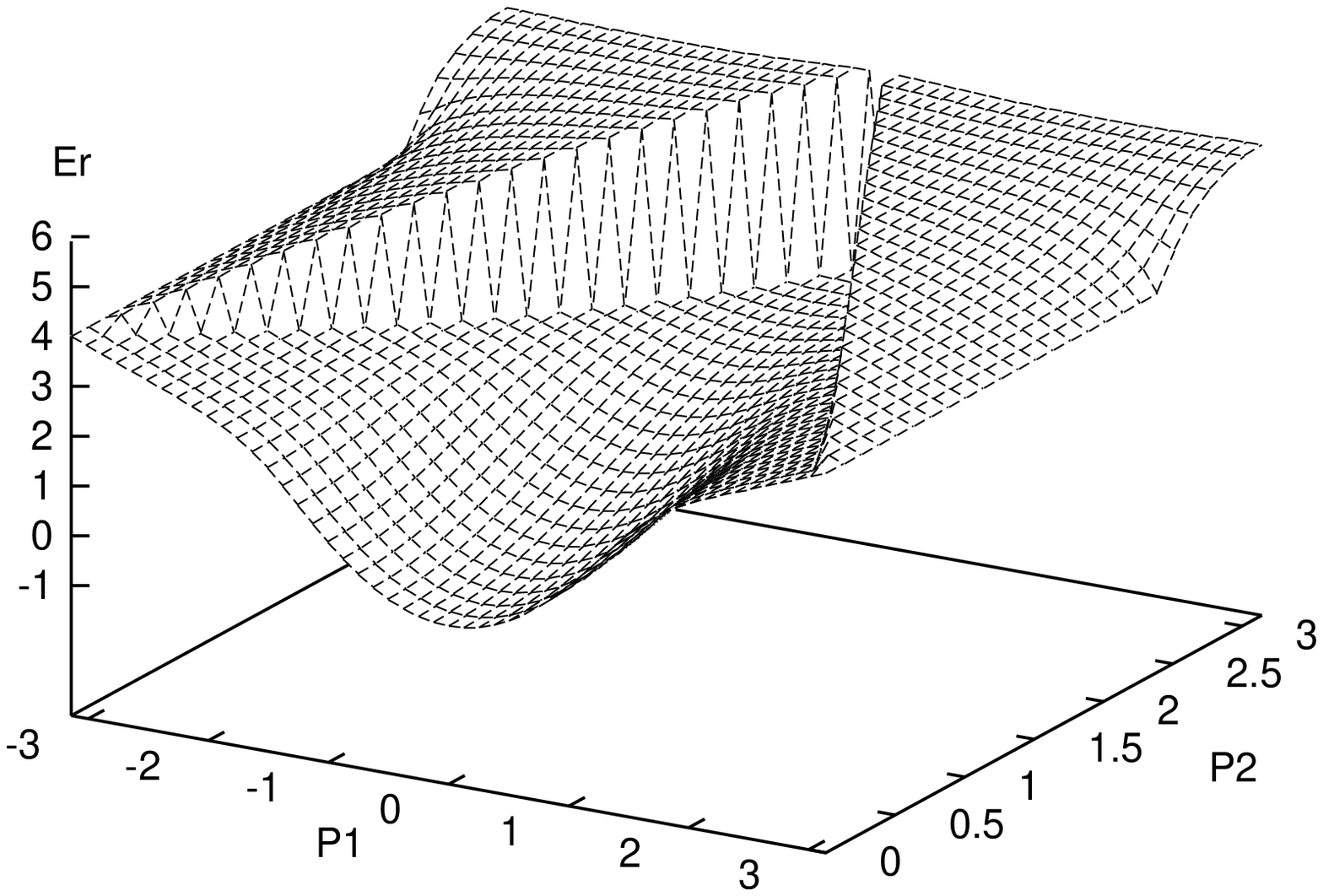}
&\includegraphics[width=8 cm] {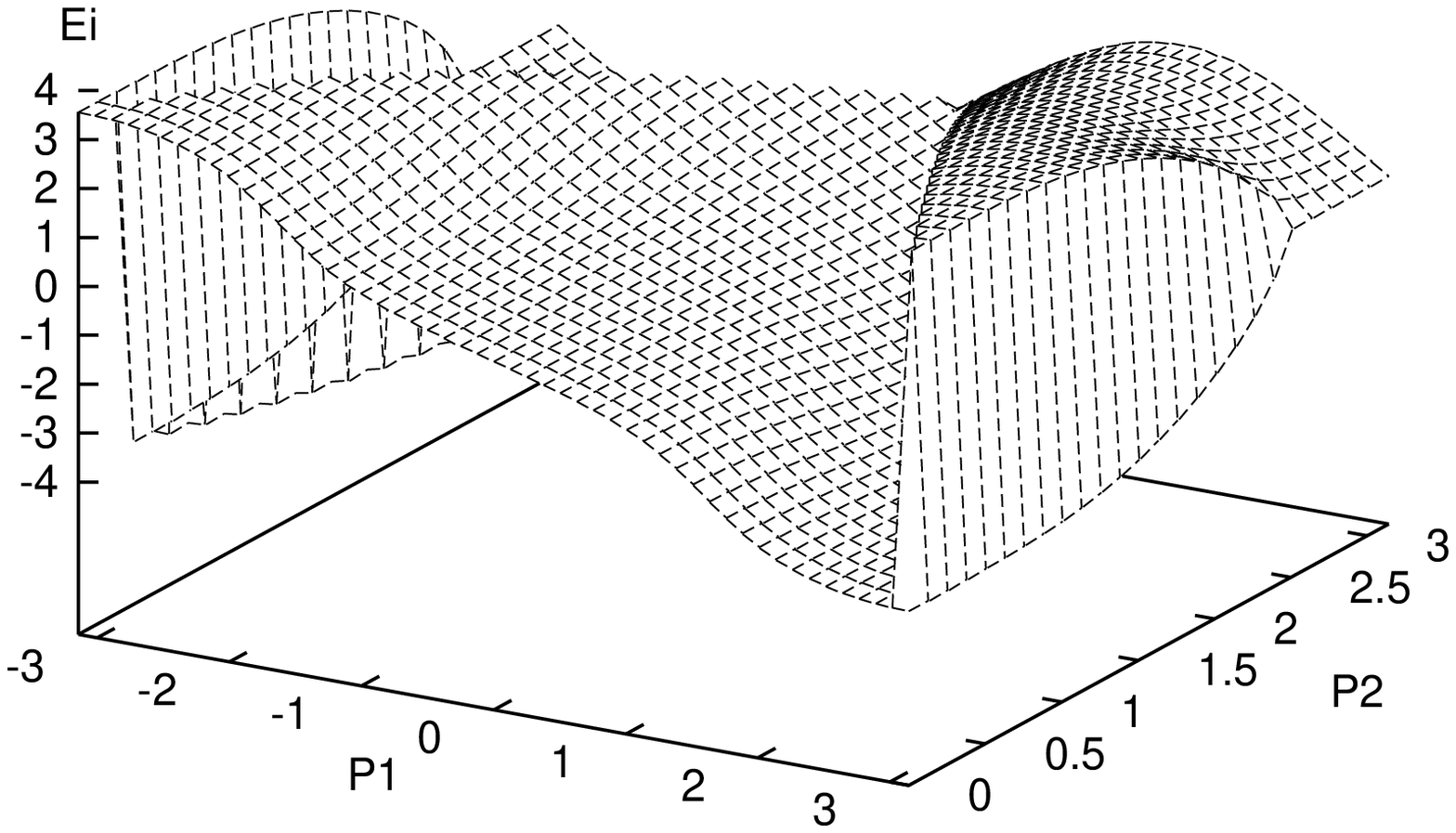}\\
\includegraphics[width=8 cm] {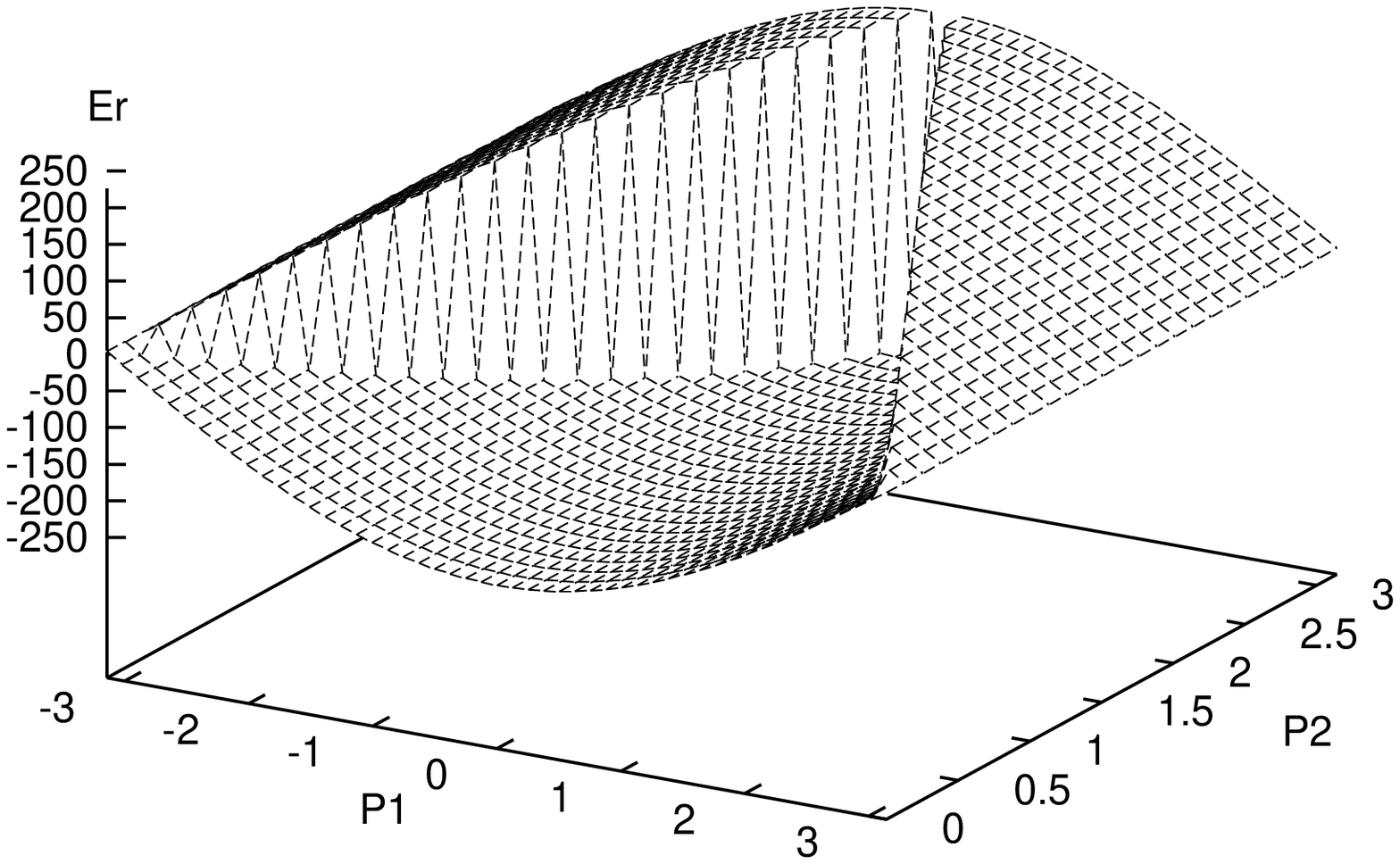}
&\includegraphics[width=8 cm] {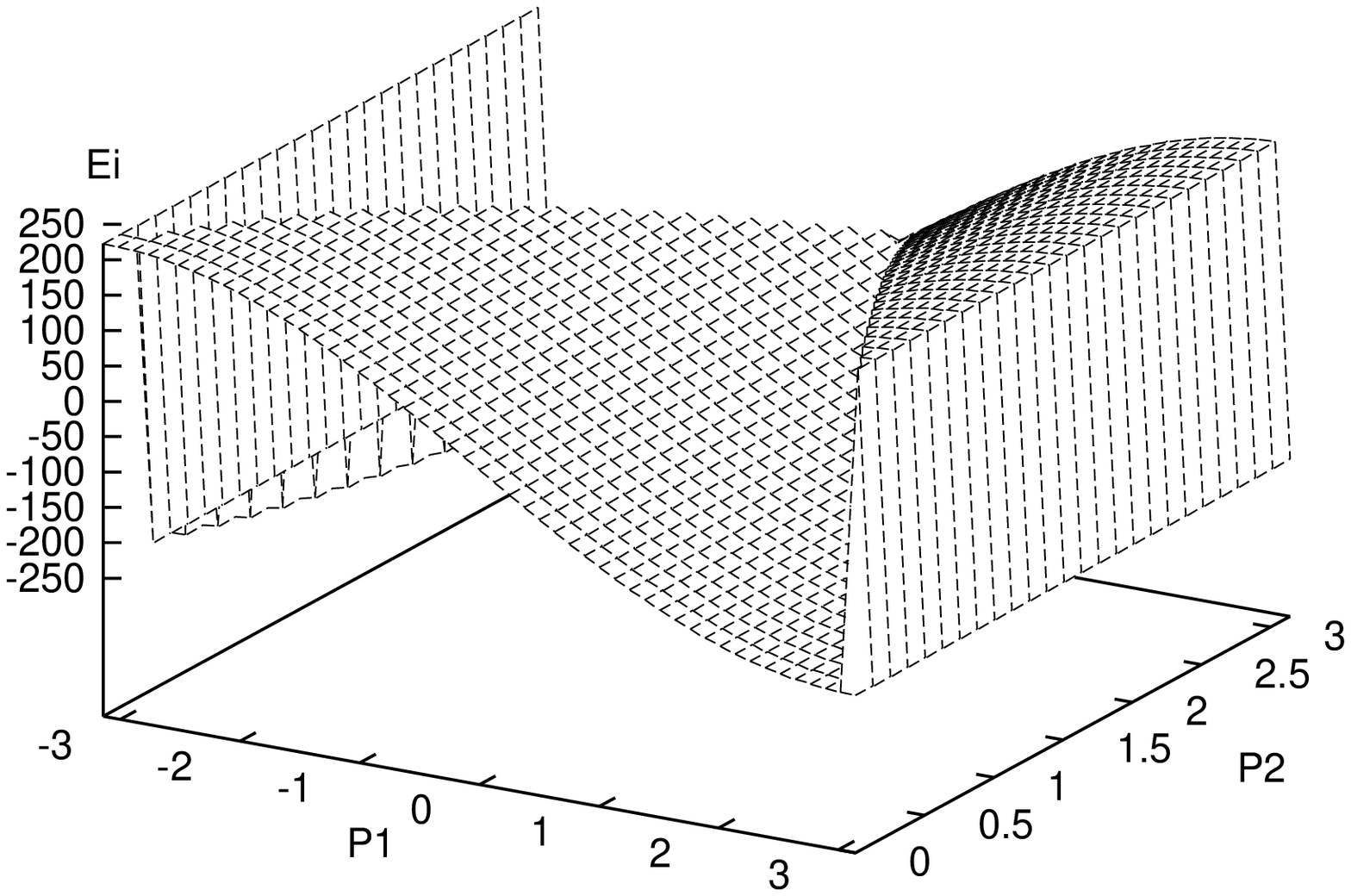}\\
\end{tabular}
\end{table}
\end{document}